\documentclass{article}
\usepackage{spconf,amsmath,graphicx}
\usepackage{makecell}
\usepackage{url}
\usepackage{enumitem}
\setlist{nosep, leftmargin=14pt}

\usepackage{mwe} 

\title{Reducing Histopathology Slide Magnification Improves the Accuracy and Speed of Ovarian Cancer Subtyping}
\twoauthors
  {Jack Breen, Nishant Ravikumar}
	{Centre for Computational Imaging and \\ Simulation  Technologies in Biomedicine, \\ School of Computing, \\ University of Leeds, UK}
  {Katie Allen, Kieran Zucker, Nicolas M. Orsi}
	{Leeds Institute of Medical Research \\ at St James’s Hospital, \\ School of Medicine, \\ University of Leeds, UK}
\begin{document}
%
\maketitle
\begin{abstract}
Artificial intelligence has found increasing use for ovarian cancer morphological subtyping from histopathology slides, but the optimal magnification for computational interpretation is unclear. Higher magnifications offer abundant cytological information, whereas lower magnifications give a broader histoarchitectural overview. Using attention-based multiple instance learning, we performed the most extensive analysis of ovarian cancer tissue magnifications to date, with data at six magnifications subjected to the same preprocessing, hyperparameter tuning, cross-validation and hold-out testing procedures. The lowest magnifications (1.25x and 2.5x) performed best in cross-validation, and intermediate magnifications (5x and 10x) performed best in hold-out testing (62\% and 61\% accuracy, respectively). Lower magnification models were also significantly faster, with the 5x model taking 5\% as long to train and 31\% as long to evaluate slides compared to 40x. This indicates that the standard usage of high magnifications for computational ovarian cancer subtyping may be unnecessary, with lower magnifications giving faster, more accurate alternatives. 
\end{abstract}
This work has been submitted to the IEEE for possible publication. Copyright may be transferred without notice, after which this version may no longer be accessible.
\section{Introduction}
\label{sec:intro}

Multiple instance learning (MIL) is commonly used for the classification of histopathology whole slide images (WSIs). In MIL methods, each WSI is represented as a bag of patches that can be separately processed to reduce the computational requirements of whole slide classification. In attention-based multiple instance learning (ABMIL) \cite{Ilse2018}, each patch is assigned a trainable attention score, with the attention-weighted average of patch embeddings used as a WSI embedding for classification through a fully connected network.  

Researchers often use different tissue magnifications in patch-based models, with the best magnification differing for subtyping different cancer types \cite{Damato2022}. Higher magnifications provide more cellular-level detail, whereas lower magnifications give more architectural context at the tissue level. Previous convolutional neural networks for ovarian cancer subtyping \cite{Breen2023review} have used tissue at 10x \cite{Farahani2022}, 20x \cite{Farahani2022,Boschman2022,Breen2023}, and 40x magnification \cite{Levine2020}. Only one previous study has directly compared results from tissue at different magnifications, reporting that performance was slightly better at 5x compared to 10x or 20x for three MIL models (including ABMIL) using a fixed set of hyperparameters \cite{Mirabadi2023}. We present the most extensive analysis of the effects of different tissue magnifications in ovarian cancer subtyping to date, including magnifications from 1.25x to 40x, with hyperparameters tuned separately at each magnification and performance evaluated on both a 5-fold cross-validation and an external test set.  


\section{Materials and Methods}
\label{sec:methods}

\begin{table}[h]
\begin{center}
\begin{tabular}{|c|c|c|}
\hline
\textbf{\makecell{Carcinoma \\ Subtype}}                     & \textbf{\makecell{Training WSIs \\ (Patients)}}       & \textbf{\makecell{Hold-out WSIs \\ (Patients)}}      \\ \hline 
\makecell{High-Grade Serous \\ (HGSC)}    & 484 (107)                      & 20 (7)                        \\ \hline
\makecell{Low-Grade Serous \\ (LGSC)}     & 23 (5)                         & 20 (6)                        \\ \hline
\makecell{Clear Cell \\ (CCC)}            & 156 (33)                       & 20 (7)                        \\ \hline
\makecell{Endometrioid \\ (EC)}           & 205 (36)                       & 20 (5)                        \\ \hline
\makecell{Mucinous \\ (MC)}               & 95 (20)                        & 20 (5)                        \\ \hline \hline
Total & 963 (201) & 100 (30) \\ \hline 
\end{tabular}%
\caption{Dataset breakdown by ovarian carcinoma subtype. Numbers in brackets indicate the number of patients.}
\label{table:dataset}
\end{center}
\end{table}

\begin{figure*}[h]

  \centering
\includegraphics[width=15.25cm]{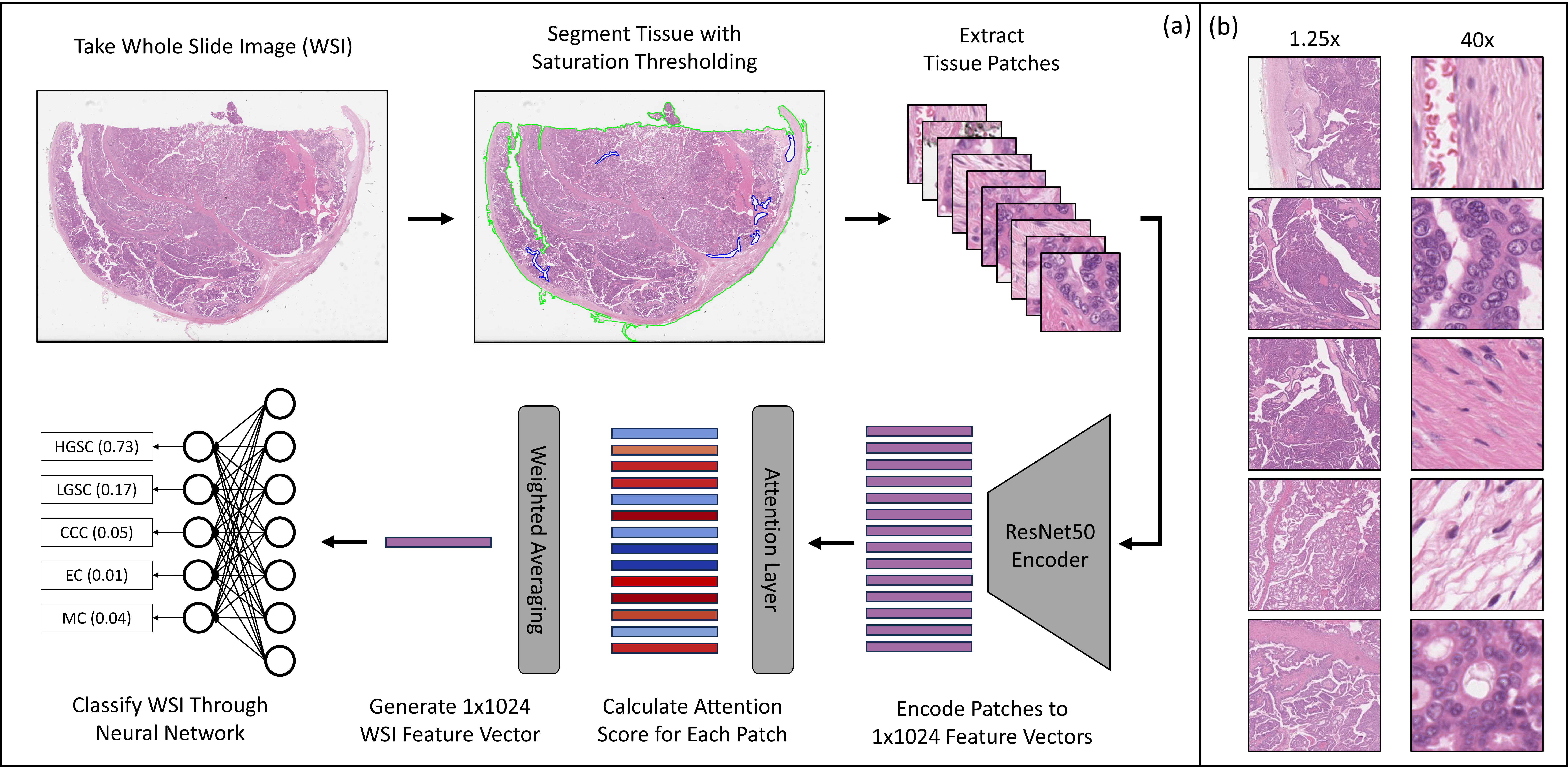}
%
\caption{\textbf{(a)} Attention-based multiple instance learning (ABMIL) \cite{Ilse2018} model pipeline for ovarian cancer subtyping. \textbf{(b)} Examples of patches from the WSI in panel (a) at the smallest (1.25x) and largest (40x) magnifications.}
\label{fig:abmil}
\end{figure*}

Our training dataset comprised 963 primary resection whole slide images (WSIs) from 201 ovarian cancer patients at Leeds Teaching Hospitals NHS Trust. This set was collated by a pathologist (KA) and all originally reported diagnoses (by a specialist gynaecological pathologist) were independently reviewed, and any cases with discrepant diagnoses were excluded. For each case, representative diagnostic tumour slides of formalin-fixed, paraffin-embedded (FFPE) haematoxylin and eosin (H\&E) stained adnexal tissue were selected. Any mounting artifacts were corrected, and slides were cleaned before being digitised at 40x magnification using a single Leica AT2 scanner. An independent set was collected following the same protocol, from which 20 WSIs of each carcinoma subtype were taken to form a balanced hold-out test set. The number of slides and patients per subtype is shown in Table \ref{table:dataset}.  

To compare the effects of different magnifications on subtyping, we used the same model training protocol (Figure \ref{fig:abmil}) for the original 40x WSIs as well as for downsampled WSIs at 20x, 10x, 5x, 2.5x, and 1.25x magnification. Before modelling, a basic saturation thresholding approach was used to separate tissue from plain background. Then, non-overlapping tissue patches were extracted such that after any downsampling the resulting patch size would be 256x256 pixels (requiring 8192x8192 pixel patches for the 1.25x experiments, for example). On average there were 68,913 patches per slide for 40x experiments, compared to only 81 for 1.25x experiments. After downsampling, an ImageNet-pretrained ResNet50 encoder was used to extract 1x1024 feature vectors from the 256x256 pixel patches. These feature vectors were finally used to train ABMIL models \cite{Ilse2018} for ovarian cancer subtyping.  


Hyperparameters were tuned at each magnification on the training dataset using an iterative strategy in which up to three hyperparameters were adjusted at any one time, with the best of these used as the baseline for following iterations. Each configuration was used to train a model for each of the five folds, with the minimum average balanced cross-entropy loss on the validation sets indicating the best model. The best overall configuration from hyperparameter tuning was used to train a model for each of the five folds, which was evaluated on the 5-fold cross-validation test sets, and on the hold-out test set by taking the average of the predictions across the five models. Model performance was evaluated using balanced accuracy, F1 score, and the area under the receiver operating characteristic curve (AUC), and was reported as the mean of 10,000 iterations of bootstrapping. Efficiency was evaluated in terms of the training and testing time on a high-performance computer (HPC), and the testing time on a personal computer (PC). Training timing did not include the time required to preprocess WSIs and extract features, but testing timing did. The HPC (used for all experiments except for the PC timing) had an NVIDIA A100 GPU and 32 AMD EPYC7742 CPUs @3.4GHz. The PC had an NVIDIA GTX 1660 GPU and a single Intel i5-4460CPU @ 3.2GHz.

Most hyperparameters affected the Adam optimizer, including the learning rate, weight decay, first moment decay ($\beta_1$), second moment decay ($\beta_2$), and stability parameter ($\epsilon$). Hyperparameters were also used to adjust the model size (specifically the attention layer and subsequent fully connected layer), the proportion of parameters dropped out before the fully connected layer during training, and the maximum number of patches randomly sampled per slide during training. Patch sampling is akin to applying dropout to the data rather than the model parameters, and thus can help to reduce overfitting \cite{Breen2023}. Approximately 80 total configurations were evaluated for each slide magnification.

\section{Results}
\label{sec:results}

\begin{table}[h]
    \centering
    \begin{tabular}{|c||c|c|c|} \hline 
         \textbf{Magnif.}&  \textbf{Balanced Accuracy}&  \textbf{AUC}& \textbf{F1 Score}\\ \hline 
         40x& 51.3\% & 0.825 & 0.516\\ \hline 
         20x& 50.6\% & 0.846 & 0.506 \\ \hline 
         10x& 52.3\% & 0.800 & 0.515 \\ \hline 
         5x&  54.0\% & 0.817 & 0.538 \\ \hline
         2.5x& \textbf{55.6\%} & 0.877 & 0.557 \\ \hline
         1.25x& \textbf{55.6\%} & \textbf{0.888} & \textbf{0.558} \\ \hline
    \end{tabular}
    \caption{Internal cross-validation results. Best results in \textbf{bold}.}
    \label{table:internalresults}
\end{table}

The optimal hyperparameters found through hyperparameter tuning and extended results are shown at \url{https://github.com/scjjb/Ovarian_Subtype_Mags}. 
The clearest trend was that the optimal number of patches used in training at lower magnifications typically covered a lower proportion of the entire slide, with the 7 patches at 1.25x covering only 9\% of an average slide, compared to the 50,000 patches at 40x covering 73\% of an average slide. The size of the optimal model weakly corresponded to the magnification, with the largest model (795,408 parameters) only used for the 40x data, and the smallest (141,424 parameters) only used for the 5x data. For the other hyperparameters, there was no obvious trend across the magnifications.

\begin{table}[b]
    \centering
    \begin{tabular}{|c||c|c|c|c|c|} \hline 
         \textbf{Magnif.}&  \textbf{HGSC}&  \textbf{LGSC}&  \textbf{CCC}&  \textbf{EC}& \textbf{MC}\\ \hline 
         40x& 0.844 & 0.000 & 0.667 & 0.668 & 0.407\\ \hline 
         20x& 0.823 & 0.000 & 0.645 & 0.668 & 0.400 \\ \hline 
         10x& 0.846 & 0.000 & 0.712 & 0.669 & 0.351 \\ \hline 
         5x& 0.853 & 0.000 & 0.764 & 0.651 & 0.423\\ \hline 
         2.5x& 0.852 & 0.000 & 0.738 & 0.682 & 0.518\\ \hline
         1.25x&  0.829 & 0.053 & 0.722 & 0.674 & 0.519 \\ \hline
    \end{tabular}
    \caption{Class-wise F1 scores in the 5-fold cross-validation.}
    \label{table:f1scores}
\end{table}

The most accurate model in cross-validation (Table \ref{table:internalresults}) was the 1.25x model, with 55.6\% balanced accuracy, 0.888 AUC, and 0.558 F1 score. Performance did not vary greatly between magnifications, with the lowest balanced accuracy being 50.6\% (at 20x), the lowest AUC being 0.800 (at 10x), and the lowest F1 being 0.506 (at 20x). Performance varied drastically between classes (Table \ref{table:f1scores}), with the least common class almost never classified correctly, whereas the most common class had an F1-score of at least 0.823 at all magnifications. The most accurate model in the hold-out test set (Table \ref{table:holdoutresults}) was the 10x model, with 62.0\% balanced accuracy, 0.850 AUC, and 0.549 F1 score. The 5x model performed similarly and had a slightly higher AUC of 0.858.

\begin{table}
    \centering
    \begin{tabular}{|c||c|c|c|} \hline 
         \textbf{Magnif.}&  \textbf{Balanced Accuracy}&  \textbf{AUC}& \textbf{F1 Score}\\ \hline 
         40x& 54.0\% & \textbf{0.860} & 0.477 \\ \hline 
         20x& 55.0\% & 0.829 & 0.485 \\ \hline 
         10x& \textbf{62.0\%} & 0.850 & \textbf{0.549} \\ \hline 
         5x& 61.0\% & 0.858 & 0.545 \\ \hline
         2.5x& 58.1\% & 0.857 & 0.516 \\ \hline
         1.25x& 58.0\% & 0.855 & 0.529 \\ \hline
    \end{tabular}
    \caption{External validation results. Best results in \textbf{bold}.}
    \label{table:holdoutresults}
\end{table}


Results from efficiency evaluations are shown in Table \ref{table:efficiency}. The fastest models to train were the 5x (10m 37s) and 1.25x (10m 49s) models. Training was always faster at lower magnifications, with the exception of the 5x model which had fewer parameters than the 2.5x or 1.25x models, resulting in faster training. The fastest models during testing were the 10x model (1m 9s per slide on HPC, 3m 8s on PC), and the 5x model (1m 11s per slide on HPC, 3m 15s on PC). The 40x model was the slowest by a wide margin for both training (3h 45m) and testing (3m 51s per slide on HPC, 7m 29s on PC). Higher magnifications took longer due to containing more patches, and lower magnifications took longer due to the large size of patches before downsampling, which took longer to load. The large patch size would not be a factor for slides with a lower native magnification, where we expect that lower magnifications would always be faster to evaluate. 

\begin{table}[h]
    \centering
    \begin{tabular}{|c||c|c|c|} \hline 
         \textbf{Magnif.}&  \textbf{\makecell{Training  \\Time   \\ Per Fold}}& \textbf{ \makecell{Testing \\ Time Per \\ Slide (HPC)}}&  \textbf{\makecell{Testing \\ Time Per\\ Slide (PC)}} \\ \hline 
         40x& 3h 45m 13s & 3m 51s & 7m 29s \\ \hline 
         20x& 44m 23s & 1m 20s & 3m 56s \\ \hline 
         10x& 12m 51s & \textbf{1m 9s} & \textbf{3m 8s} \\ \hline 
         5x& \textbf{10m 37s} & 1m 11s & 3m 15s \\ \hline 
         2.5x& 12m 29s & 1m 24s & 4m 28s \\ \hline 
         1.25x& 10m 49s & 1m 58s & 5m 42s \\ \hline
    \end{tabular}
    \caption{Average training and testing times at different magnifications. Test times (preprocessing and classification) were analysed using a balanced subset of 20 WSIs from the hold-out test set, on a high-performance computer (HPC) and a personal computer (PC). Fastest indicated in \textbf{bold}.}
    \label{table:efficiency}
\end{table}


\section{Discussion}
\label{sec:discussion}

Our results indicate that the standard 40x and 20x magnifications used in the clinical setting may not be best for computational ovarian cancer subtyping, with 10x and 5x each giving higher accuracy in cross-validation and an independent test set. This reflects previous results where 5x and 10x MIL models were found to outperform 20x \cite{Mirabadi2023}. The balanced accuracies and F1 scores reported in this study are lower than those in some other studies despite using similar methodologies \cite{Farahani2022,Levine2020,Mirabadi2023}, likely due to the few LGSC cases leading to very poor performance at classifying this specific cancer subtype. 

This study was limited to a single model type (ABMIL) using WSIs at a single magnification. Different models, such as graphs and transformers, may perform differently across magnifications due to their ability to model relationships between patches. Multi-magnification models may be able to improve performance by combining the cellular and histoarchitectural information at different magnifications, akin to how pathologists would interpret different morphological features to obtain overall diagnostic insight from multiple different magnifications of a single slide. This study is also limited to data from a single data centre, so it is unclear how model generalisability is affected by tissue magnification. 

While the routine collection of data at lower magnifications would offer efficiency benefits (reduced storage requirements and cost), this is not clinically viable as histopathologists require higher magnifications for manual review. The benefit of using lower magnifications in modelling (aside from any improvement in classification accuracy) is that it reduces the computational requirements, allowing models to be deployed directly to the pathology lab instead of requiring WSIs to be sent off-site at additional expense. Such models may be used as assistive tools to improve the accuracy and efficiency of the pathological workflow, reducing the need to outsource difficult cases to gynaecological experts. 






\section{Conclusion}
In this paper, we presented the most extensive evaluation of tissue magnifications for ovarian cancer subtyping to date. We tuned attention-based multiple instance learning models at six different magnifications from 1.25x to 40x. We found that reducing magnification from the standard 40x or 20x did not degrade performance, and in many cases, improved performance. Specifically, 10x and 5x magnifications gave the best balanced accuracy in external validation while also being the fastest models to run. The performance of even lower magnification models was also impressive, with 1.25x magnification outperforming 40x and 20x by most metrics. 

\section{Acknowledgments}
\label{sec:acknowledgments}
JB is supported by the UKRI Engineering and Physical Sciences Research Council (EPSRC) [EP/S024336/1]. NMO receives research funding from 4D Path. All other authors declare no conflicts of interest. For the purpose of open access, the author has applied a Creative Commons Attribution (CC BY) licence to any Author Accepted Manuscript version arising from this submission. All code is provided at \url{https://github.com/scjjb/Ovarian_Subtype_Mags}, which is an extension of the original CLAM pipeline \cite{Lu2021}.

\section{Compliance with Ethical Standards}
\label{sec:Compliance}
This study was conducted retrospectively using human subject data and received approval from the Wales Research Ethics Committee [18/WA/0222] and the Confidentiality Advisory Group [18/CAG/0124].

\bibliographystyle{IEEEbib}
\bibliography{main}

\begin{thebibliography}{1}

\bibitem{Ilse2018}
Maximilian Ilse, Jakub Tomczak, and Max Welling,
\newblock ``Attention-based deep multiple instance learning,''
\newblock in {\em International conference on machine learning}. PMLR, 2018, pp. 2127--2136.

\bibitem{Damato2022}
Marina D'Amato, Przemys{\l}aw Szostak, and Benjamin Torben-Nielsen,
\newblock ``A comparison between single-and multi-scale approaches for classification of histopathology images,''
\newblock {\em Frontiers in Public Health}, vol. 10, pp. 892658, 2022.

\bibitem{Breen2023review}
Jack Breen, Katie Allen, Kieran Zucker, Pratik Adusumilli, Andrew Scarsbrook, Geoff Hall, Nicolas~M Orsi, and Nishant Ravikumar,
\newblock ``Artificial intelligence in ovarian cancer histopathology: a systematic review,''
\newblock {\em NPJ Precision Oncology}, vol. 7, no. 1, pp. 83, 2023.

\bibitem{Farahani2022}
Hossein Farahani, Jeffrey Boschman, David Farnell, Amirali Darbandsari, Allen Zhang, Pouya Ahmadvand, Steven~JM Jones, David Huntsman, Martin K{\"o}bel, C~Blake Gilks, et~al.,
\newblock ``Deep learning-based histotype diagnosis of ovarian carcinoma whole-slide pathology images,''
\newblock {\em Modern Pathology}, vol. 35, no. 12, pp. 1983--1990, 2022.

\bibitem{Boschman2022}
Jeffrey Boschman, Hossein Farahani, Amirali Darbandsari, Pouya Ahmadvand, Ashley Van~Spankeren, David Farnell, Adrian~B Levine, Julia~R Naso, Andrew Churg, Steven~JM Jones, et~al.,
\newblock ``The utility of color normalization for ai-based diagnosis of hematoxylin and eosin-stained pathology images,''
\newblock {\em The Journal of Pathology}, vol. 256, no. 1, pp. 15--24, 2022.

\bibitem{Breen2023}
Jack Breen, Katie Allen, Kieran Zucker, Geoff Hall, Nishant Ravikumar, and Nicolas~M Orsi,
\newblock ``Predicting ovarian cancer treatment response in histopathology using hierarchical vision transformers and multiple instance learning,''
\newblock {\em arXiv preprint arXiv:2310.12866}, 2023.

\bibitem{Levine2020}
Adrian~B Levine, Jason Peng, David Farnell, Mitchell Nursey, Yiping Wang, Julia~R Naso, Hezhen Ren, Hossein Farahani, Colin Chen, Derek Chiu, et~al.,
\newblock ``Synthesis of diagnostic quality cancer pathology images by generative adversarial networks,''
\newblock {\em The Journal of pathology}, vol. 252, no. 2, pp. 178--188, 2020.

\bibitem{Mirabadi2023}
Ali~Khajegili Mirabadi and Graham Archibald,
\newblock ``Heram: Multi-magnification graph-structured whole slide image representation,''
\newblock 2022.

\bibitem{Lu2021}
Ming~Y Lu, Drew~FK Williamson, Tiffany~Y Chen, Richard~J Chen, Matteo Barbieri, and Faisal Mahmood,
\newblock ``Data-efficient and weakly supervised computational pathology on whole-slide images,''
\newblock {\em Nature biomedical engineering}, vol. 5, no. 6, pp. 555--570, 2021.

\end{thebibliography}

\end{document}